\documentclass[a4paper,fleqn,usenatbib]{mnras}

\usepackage{newtxtext,newtxmath}

\usepackage[T1]{fontenc}
\usepackage{ae,aecompl}

\usepackage{graphicx}	
\usepackage{amsmath}	
\usepackage{amssymb}	
\usepackage{graphics}

\usepackage{xcolor}

\title[Determining the rotation count of pulsars]{An algorithm for determining the rotation count of pulsars}

\author[Paulo C. C. Freire]{
Paulo C. C. Freire$^{1}$\thanks{E-mail: pfreire@mpifr-bonn.mpg.de}
and Alessandro Ridolfi$^{1}$
\\
$^{1}$Max-Planck-Institut f\"{u}r Radioastronomie, Auf dem H\"{u}gel 69, D-53121 Bonn, Germany\\
}

\date{Accepted XXX. Received YYY; in original form ZZZ}

\pubyear{2015}

\begin{document}
\label{firstpage}
\pagerange{\pageref{firstpage}--\pageref{lastpage}}
\maketitle

\begin{abstract}
We present here a simple, systematic method for determining the
correct global rotation count of a radio pulsar; an essential
step for the derivation of an accurate phase-coherent ephemeris.
We then build on this method by developing
a new algorithm for determining
the global rotational count for pulsars with sparse timing data sets.
This makes it possible to obtain phase-coherent ephemerides for pulsars
for which this has been impossible until now. As an example, we do this
for PSR~J0024$-$7205aa, an extremely faint MSP
recently discovered in the globular cluster 47~Tucanae.
This algorithm has the potential to significantly reduce the number of observations and the amount of telescope time 
needed to follow up on new pulsar discoveries.
\end{abstract}

\begin{keywords}
methods: data analysis -- (stars:) pulsars: general -- (stars:) pulsars: individual (PSR J0024$-$7205aa)
\end{keywords}



\section{Introduction}

In the 50 years since the discovery of radio pulsars, their study has been a major
scientific success story. They are extremely versatile and powerful tools for studying
fundamental physics, in particular the study of gravity, the fundamental properties of
space-time and of gravitational waves
(see \citealt{wex14,bbc+15} and the many references therein).
Measurements of large neutron star masses (e.g., \citealt{afw+13,fpe+16}) 
had profound implications for the study of cold nuclear matter at supra-nuclear
densities (see \citealt{of16} and references therein), a fundamental question in nuclear physics.
The orbital properties and abundance of double neutron star systems
provided the first estimates of the rate of NS-NS mergers, the first ``guaranteed'' source
of events for ground-based gravitational wave detectors. Furthermore pulsars can be
used to detect very low frequency gravitational waves
directly \citep{abb+15,dcl+16,rhc+16,vlh+16}.

Apart from these ``fundamental physics'' applications, radio pulsars are also superb
astrophysical tools: they allow a much deeper understanding of the late stages
of the evolution of massive stars, including the supernova events
(important for understanding the origin of the most useful pulsars, the recycled pulsars, see e.g.,
\citealt{lor08,tkf+17}), the dynamics and history of globular clusters
and the ionized interstellar medium.

\subsection{Pulsar timing}

Most of these applications rely on a single, simple technique,
{\em pulsar timing}. This technique achieves its full power for
recycled pulsars. Put simply, this consists
of the study of the times of arrival (TOAs) of the pulses
(normally determined by adding many individual pulses coherently in phase)
at one or several telescopes, most often radio telescopes.
These TOAs can be determined very accurately, in some cases
with precisions better than 100 ns, but more commonly a few
$\mu$s; they correspond to times where a particular longitude
of the pulsar (normally close to the spin phase
where the observed radio emission is at a maximum) is aligned towards the Earth.
Therefore, in between any two radio pulses the pulsar has rotated an integer number of times. 

A good timing solution (henceforth a {\em phase-coherent ephemeris}) must be
capable of accurately predicting the TOAs. This ephemeris consists of
a specified mathematical description with a few crucial free parameters that describe
the spin of the pulsar plus the transformation between the reference frame of the
pulsar and that of the receiving telescope. In the reference frame of most
recycled pulsars, their rotation can be described by a spin frequency ($\nu$),
plus a small, but constant and negative spin frequency derivative
($\dot{\nu}$). In the reference frame of the receiving radio telescope, these
arrival times are affected by the motion of the pulsar relative to the radio
telescope. This has several components: the observatory's motion caused by
the Earth's rotation, the motion of the Earth relative to
the Solar System barycentre (SSB), and the motion of the pulsar relative
to the SSB, which might be affected by the pulsar's own orbital motion (if it happens to have
any companion(s)). In order to correct for the movement of the radio telescope
relative to the SSB, we need to have an estimate
of the position of the pulsar in the sky; i.e., the right ascension ($\alpha$),
declination ($\delta$) and, for measurements spanning many years, the proper motion in these
two coordinates ($\mu_{\alpha}, \mu_{\delta}$). Finally, if the pulsar is in a binary system, then its
orbital motion can be normally parameterized by 5 Keplerian parameters: in the
\cite{dd86} model (known as the DD model), for example,
these parameters are the orbital period ($P_b$), the semi-major axis of the pulsar's
orbit projected along the line of sight ($x$), the orbital eccentricity ($e$),
the longitude of periastron ($\omega$) and the time of passage through periastron ($T_0$). In some
cases, a few additional ``post-Keplerian" parameters can be measured, these are caused by geometric
\citep{kop96} and relativistic (e.g., \citealt{dt92}) effects.
In the case of ``black widow'' or ``redback'' pulsars, where there are
Newtonian perturbations to the orbit, additional parameters might
be necessary (e.g., \citealt{2016MNRAS.462.1029S}).
It is clear from the number of parameters (and their measurement precision) that these
phase-coherent ephemerides provide a wealth of scientific information.

These ephemerides are derived from the TOAs using a timing
program, like {\sc tempo}\footnote{\url{http://tempo.sourceforge.net/}},
{\sc tempo2} \citep{2006MNRAS.369..655H,2006MNRAS.372.1549E} or 
{\sc PINT}\footnote{\url{https://github.com/nanograv/PINT}}; in what follows
we will be using {\sc tempo} because it allows a very simple implementation of the methods to be described.
For each TOA, {\sc tempo} will first correct it using tables where the local time standard at the
radio telescope is compared to a more stable time scale, e.g., the universal coordinated time
(UTC). Then, the position of the radio telescope relative to the SSB is calculated, first by using
Earth rotation tables to calculate where the radio telescope is relative to the Earth's
centre, then by using a Solar System ephemeris (like DE 421, \citealt{fwb08} or DE 430, \citealt{fwb+14})
to translate the latter to the SSB. This vector is then projected along the direction to
the pulsar (derived from $\alpha,\delta,\mu_{\alpha}$ and $\mu_{\delta}$) to calculate the effect of
the motion of the radio telescope relative to the SSB on the TOAs. Then, if the pulsar
is in a binary, a binary model (like the aformentioned DD model)
is used to subtract the time delays caused by the orbital motion.
It is only after this stage that {\sc tempo} calculates the {\em residuals}: these
are the TOAs minus the prediction of the model {\em for the corresponding rotation of the pulsar}.
The best-fit timing parameters are determined by varying the timing parameters
in a way that minimizes the sum of the squares of the residuals.

\subsection{The problem}

In the last paragraph, we have placed the emphasis on one fundamental point, which
is rarely discussed in the literature: {\sc tempo} can only determine a
phase-coherent ephemeris if it is
comparing a TOA with the model estimate for that pulse, not the one before or
after. This can only be achieved {\em after} the correct rotation count between
any two TOAs (a set of integers) has been established. This is what
we call a {\em global rotation count}.

As we will see, when a new pulsar is discovered, the rotation count within a
single observation is generally well known. However, the rotation count between
successive observations is generally {\em not} known.
As we will show, this issue is easy to solve for most pulsars, but it is hard to solve if
there are no closely spaced detections of the pulsar, as in the case of
scintillating pulsars, or pulsars in eclipsing systems. In the case
of very faint pulsars, the timing precision might be too poor for the determination
of the rotation count even for closely spaced observations.

The number of known faint pulsars
has been growing significantly, but the time available at the major radio telescopes that
can detect them remains unchanged,
this implies automatically less time available to follow up each new pulsar. This can be
partly compensated by more sensitivity and larger bandwidths, but also a more careful
coordination of observations, and by dropping many pulsars that are scientifically less
rewarding.

However, as we will see below, even if a pulsar is seen to have a high scientific
value and its observation becomes a priority at a particular radio telescope,
finding the correct rotation count might still be extremely difficult to achieve.
Frequent eclipses or the effects of scintillation mean that even a dense set of
observations does not necessarily translate into a dense set of detections with a minimal useful signal-to-noise ratio (S/N).
Thus, in some cases, a phase-coherent ephemeris has not been obtained to date.

\subsection{Structure of the paper}

In Section~\ref{sec:observations} we present the observations and measurements
of TOAs for a
recently discovered pulsar, PSR~J0024$-$7205aa (henceforth 47~Tuc~aa);
an example of a pulsar for which no phase-coherent ephemeris could be derived until now, despite the large number of observations available.
The TOAs from this pulsar are then used to illustrate the concepts discussed in this
paper, always using {\sc tempo} as a practical tool for deriving results: we believe that the use of a this specific program on
a specific example is helpful for the illustration of the
concepts presented in the paper.
In Section~\ref{sec:connection}, we present the concept
of global rotation count in more detail and describe a simple, standard
method for achieving it that can be applied to most pulsars; this assumes
familiarity with the {\sc tempo} timing package. This
standard method can determine global rotation counts for most pulsars but it
cannot achieve this for 47~Tuc~aa owing to the extreme sparsity of its detections.
In Section~\ref{sec:mutiple_trials}, we highlight a method for determining the
global rotation count for such pulsars with a minimal number of {\sc tempo}
iterations, and present its implementation as a UNIX shell script.
In Section~\ref{sec:47Tucaa}, we use this method to determine the rotation count of
47~Tuc~aa, presenting a previously unavailable phase-coherent ephemeris for this pulsar
and discussing the significance of its parameters.
Finally, in Section~\ref{sec:conclusions}, we summarize our results and highlight
some of the possibilities opened up by the method discussed here.

Throughout this work, we will assume basic familiarity with the concepts of
pulsar signal analysis, in particular dedispersion, folding and the use of low-noise
pulse profile templates to estimate TOAs; for a review see \cite{lk04}.
We will introduce some concepts as we go along, whenever this happens
the new concept is presented in italic.

\section{Observations and data reduction of 47 Tuc aa}
\label{sec:observations}

The pulsar discussed in this work, 47~Tuc~aa, is one of the 25 radio millisecond
pulsars (MSPs) known in the globular cluster NGC 104, also known
as 47~Tucanae (henceforth 47~Tuc). All these pulsars are detected in the same data
(the telescope beam covers all their positions simultaneously), obtained
during 519 observations of 47~Tuc carried out from 1997 until 2013 (i.e., a time span of 16 years)
with the 64-m Parkes radio telescope in New South Wales, Australia. These observations are
part of a long-term project dedicated to these pulsars. The project's scientific motivation,
setup of the observations, data taking (with the Analogue Filterbank, or AFB)
and reduction and some of its results are
described by \cite{rft+16}, \cite{frk+17} and their references.

47~Tuc~aa was discovered in this data set by \cite{phl+16}.
With a spin period of only 1.845 ms, it is the fastest-spinning pulsar
known in 47~Tuc. It also has the highest dispersion measure (DM) for any known
pulsar in the cluster, $24.941(7) \, \rm pc\, cm^{-3}$. The detections of
this object are extremely rare: in the 519 observations of the data set,
the pulsar was only detected (in the sense of
yielding at least one usable TOA) on 22 occasions; this represents a detection rate of 4.2 per cent, and an average of 1.38 detections per year.
In most of these detections the signal-to-noise ratio (S/N) is smaller than 6.
Given this extreme sparsity, no phase-coherent ephemeris could be derived by \cite{phl+16}.

The reason for the small number of detections is the small flux density of the
pulsar combined with diffractive scintillation: like most pulsars in 47~Tuc,
47~Tuc~aa has a flux density that is well
below the telescope's sensitivity limit. Most of these pulsars are only detectable
when scintillation amplifies their signal (see e.g., \citealt{clf+00}).
47~Tuc~aa is so faint that only very rarely
is the scintillation amplification large enough to make the pulsar detectable, even in
observations lasting 8 hours.
This means that although we can choose when to observe the cluster
and the pulsars in it, we cannot choose when to detect any particular pulsar.
This is also one of the reasons why, despite the lack of improvement 
of the sensitivity, new pulsars are still being discovered in this data set:
for instance, 47~Tuc~aa is not detected on the vast majority ($\sim \, 96\%$) of observations,
and other even fainter pulsars might be appearing even more infrequently.

Although no phase-coherent ephemeris was presented for 47~Tuc~aa in \cite{phl+16}, the
few detections of this object show that, after correction for the
Doppler shift due to the Earth's motion, its spin period does not change
measurably with time. This implies that, unlike most MSPs in 47~Tuc,
this particular object is not part of a binary system. Apart from this,
not much was known about it, in particular its location relative to the
centre of the cluster, its proper motion or its spin period derivative. 
These parameters are exquisitely well determined for the other 22 MSPs in
47~Tuc with phase-coherent ephemerides \citep{rft+16,frk+17}.

\section{Rotation count and how to determine it}
\label{sec:connection}

\subsection{Rotation count within an observation}
\label{sec:phase_within_observation}

\begin{figure}
\centering
	\includegraphics[width=0.6\columnwidth]{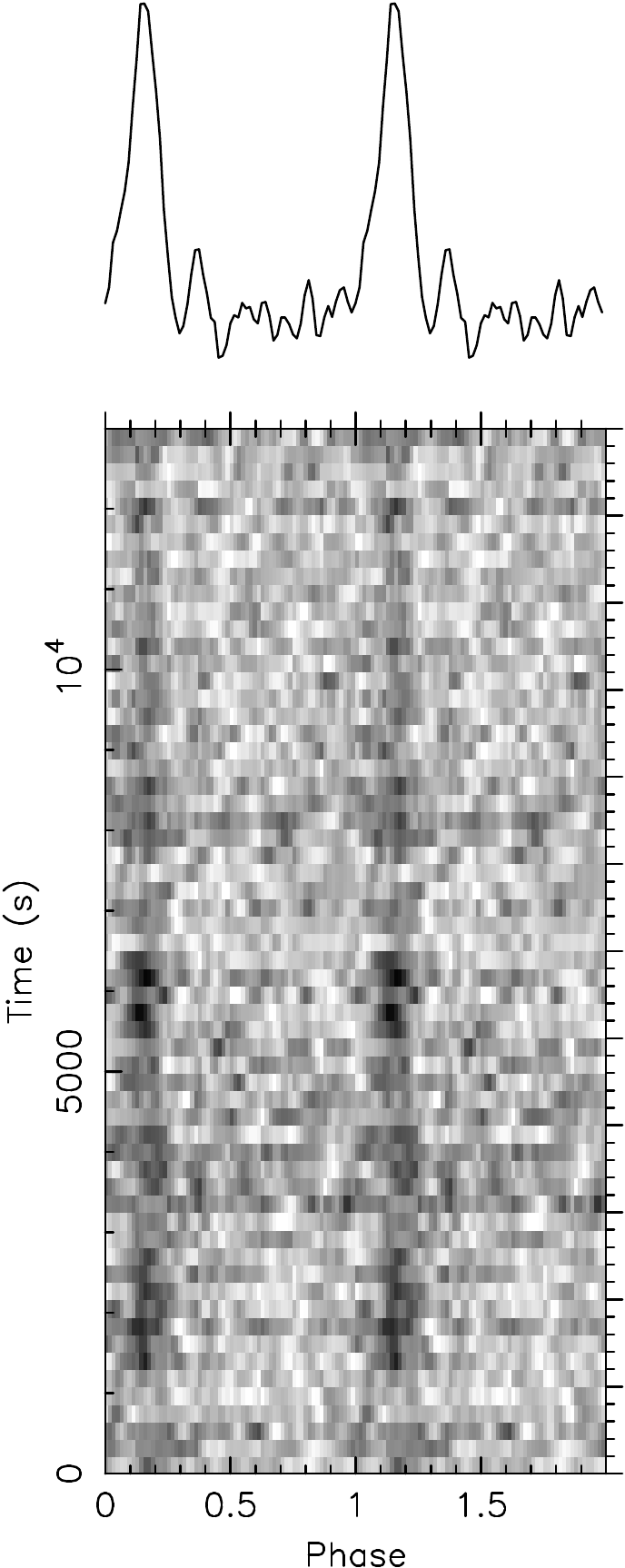}
    \caption{Best detection of 47~Tuc~aa from an observation made on 2007 August 2.
    In the main greyscale plot, we display the
    radio intensity at 20~cm (darker meaning larger intensity) as a function of the spin phase
    (displayed twice for clarity) and time, which is divided into 60 segments.
    As we can see, the pulsar signal (the vertical
    gray line) does not drift perceptibly in phase, indicating that the spin period
    used to fold these data is accurate. The top plot represents the integrated radio
    intensity (vertical axis) as a function of spin phase, also displayed twice for clarity.}
    \label{fig:observation}
\end{figure}

In Fig.~\ref{fig:observation}, we display one of the best detections of
47~Tuc~aa, from an observation on MJD = 54314
(2007 August 2). In this plot, we can see that the pulsations
appear at a constant spin phase. This shows conclusively
that the estimate of the spin frequency ($\nu\, = \, 541.893656\, $Hz),
determined from the search procedure \citep{phl+16}, is precise enough to
predict the arrival times of the pulses within that observation. 
If the spin period were in error, we should see a drift of the
phase with time.

\begin{figure}
	\includegraphics[width=\columnwidth]{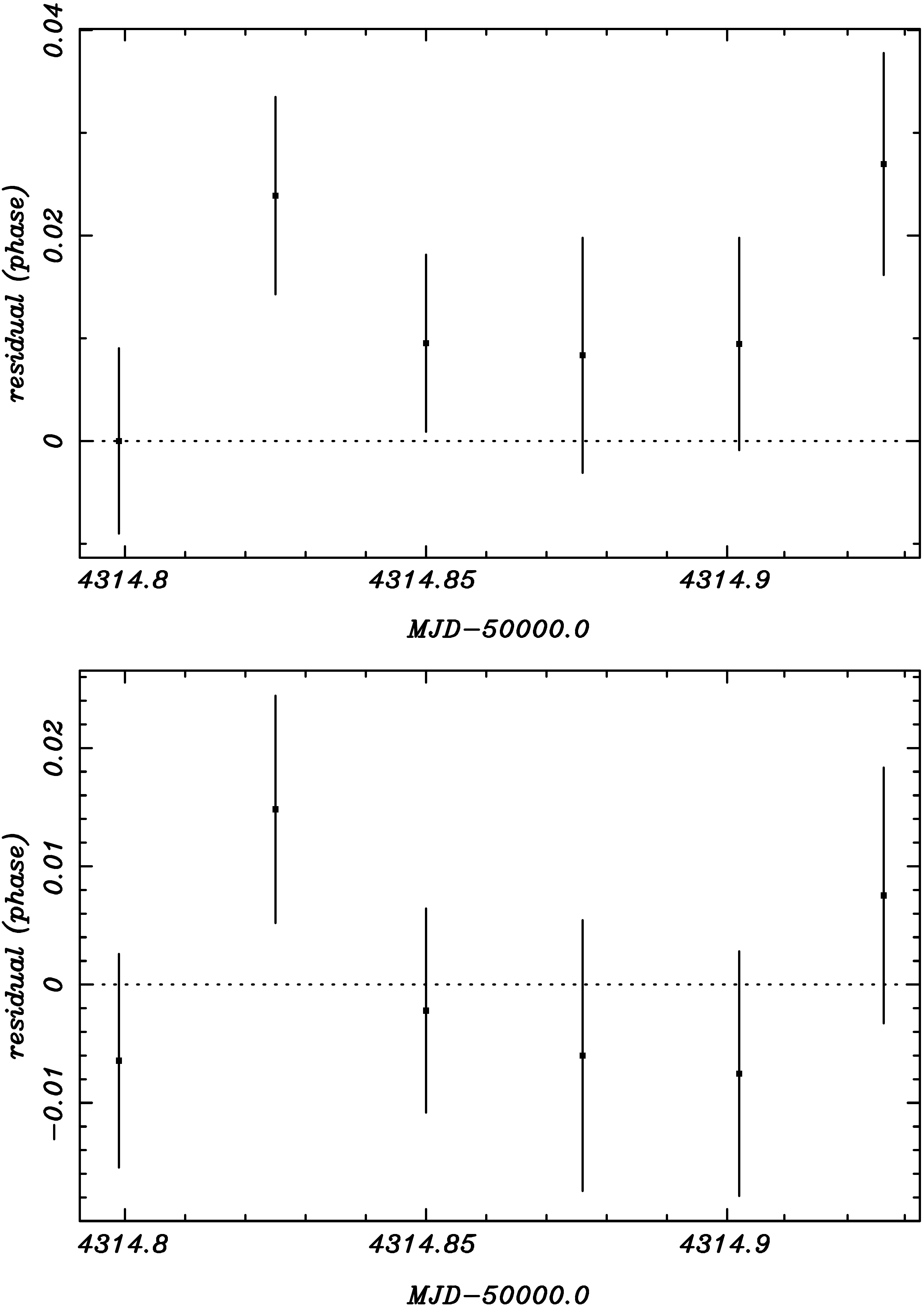}
    \caption{Pre-fit (top) and post-fit (bottom) residuals for the 6 TOAs derived from the best
    observation of 47~Tuc~aa, taken on 2007 August 2. The horizontal axis represents
    time and the vertical axis represents the residual, in spin phase units.
    0.01 of the spin phase corresponds to 18.5 $\mu$s, which is comparable to the
    TOA uncertainties.}
    \label{fig:first_residuals}
\end{figure}

This can be confirmed with {\sc tempo} in the following way:

\begin{enumerate}
\item Group the 60 sub-integrations displayed in Fig.~\ref{fig:observation}
into, for example, 6 longer sub-integrations and, using the techniques
described in \cite{rft+16}, make a single TOA for each sub-integration.
A TOA list suitable for
use in {\sc tempo} should look like
this\footnote{The program accepts a variety of formats.
In this case we chose use the ``Princeton'' format
because of its simplicity.}:

\begin{scriptsize}
\begin{verbatim}
MODE 1
7               1390.000 54314.7954878129250   16.660                 0.00000
7               1390.000 54314.8211805391308   17.735                 0.00000
7               1390.000 54314.8468749751720   15.921                 0.00000
7               1390.000 54314.8725694346731   21.125                 0.00000
7               1390.000 54314.8982638745009   19.083                 0.00000
7               1390.000 54314.9222205617143   19.950                 0.00000
\end{verbatim}
\end{scriptsize}

The ``MODE 1'' flag indicates that {\sc tempo} will make a weighted fit,
with the weight of each TOA given by the inverse of its uncertainty. 
The first column indicates the radio telescope where the data were taken (in this case
7 identifies Parkes), the second column the central radio frequency in MHz,
the third column is the TOA itself in MJD, the fourth column is the TOA uncertainty in
$\mu$s and the last one indicates any previous DM corrections (none in this case).

\item Make a simple ephemeris file (``47Tucaa.par'') containing the best
estimate of the spin frequency (we start from the estimate in
\citealt{phl+16}), no spin period derivative (this is generally very small), and
the coordinates of the centre of 47~Tuc (the pulsar is not likely
to be more than $1.2 \arcmin$ away from that centre, see \citealt{frk+17}, except
in one case, 47~Tuc~X, see \citealt{rft+16}) and the proper
motion of the cluster ({\em idem}). We also put in a parallax
derived from the known cluster distance (4.69 kpc, see discussion in
\citealt{frk+17} and references therein); this has
little effect on the timing but it is a known quantity anyway.
This ephemeris looks like this:

\begin{scriptsize}
\begin{verbatim}
PSR              J0024-7205AA
RAJ           00:24:05.67
DECJ         -72:04:52.62 
PMRA                 5.16 
PMDEC               -2.85
PX                 0.2132
F0             541.893656 1
F1                      0
PEPOCH              51600
DM                 24.971
EPHEM               DE421
CLK                 UTC(NIST)                       
UNITS               TDB
NITS 1
\end{verbatim}
\end{scriptsize}

The flag ``1'' after F0 (the spin frequency, $\nu$) means
that we are fitting for this quantity. The reference
epoch { (given by the ``PEPOCH'' flag) should generally be the barycentric
time at which $\nu$ was measured. For a MSP with a 
small variation of the spin period, this is not so crucial, 
so we set PEPOCH to MJD = 51600 (2000 February 26) because it is the
reference epoch used for all other} ephemerides in \cite{frk+17}.

With these files, we then run {\sc tempo} and look at the residuals. The pre-fit and post-fit residuals plots are displayed in Fig.~\ref{fig:first_residuals}; they have reduced $\chi^2$ of 4.24 and 1.06 respectively.
\end{enumerate}

In that Figure, {\sc tempo} confirms what was already obvious from Fig.~\ref{fig:observation}: the ephemeris ``47Tucaa.par'' predicts the correct rotation count within the observation:
all TOAs are arriving within a very small phase window (from 0.0, defined automatically
as the phase of the first TOA, to 0.03, a range that is of the same order as the TOA
uncertainties). This means that the rotation count (5933580)
between the first and last TOAs from the barycentric spin frequency estimated from
the initial spin frequency is correct, i.e., this group of TOAs is
{\em phase-connected} (henceforth {\em connected}). After fitting for $\nu$,
the situation improves further, with all residuals even closer to 0.
 
From this set of TOAs, we can already have a very precise measurement of the
spin frequency (or period, $P \equiv 1 / \nu$).
According to the {\sc tempo} output (printed in file tempo.lis)
the time elapsed between the first and last TOA is 3.04 hours,
more precisely $\delta T_{t}\, = \, $10949.709495(26) s in the telescope reference frame
and  $\delta T_{b}\, =\, $10949.713006(26) s in the SSB.
Dividing the latter by the rotation count mentioned above, we obtain a
barycentric spin period ($P$) of 0.0018453805301 s, exactly the value
returned by {\sc tempo}. The uncertainty of the spin period $\delta P$
should be the uncertainty of $\delta T_{b}$ ($26 \, \mu \rm s$) divided
by the number of rotations, i.e., $\delta P\, = \, 4.8\, \times\, 10^{-12}\, \rm s$.
The uncertainty estimated by {\sc tempo} is actually smaller
($3.7\, \times\, 10^{-12}\, \rm s$) since it does not only take into account
the first and last TOAs, but the other 4 intermediate ones as well.
The parameters from this fit are presented in row 1 of Table~\ref{tab:parameters}.

For the remainder of this paper it is important to notice that
the absolute phase of the pulses is immaterial, since {\sc tempo} automatically
assigns a spin phase of zero to the first TOA. All that matters is
the rotation count between TOAs. By minimizing the
residual rms, all that {\sc tempo} did was to adjust the spin frequency
in order to remove the {\em residual slope} from this single observation.
Such residual slopes are essential because they give us the initial constraints on the global
rotation count and the ephemeris to be derived from it.

\begin{table*}
	\centering
	\caption{Post-fit parameters for several iterations described in the text. *Parameter not being fit, derived from initial assumptions (see text).\label{tab:parameters}}
	\begin{tabular}{l l l l l c c} 
		\hline
		Iteration  & $\nu$ & $\dot{\nu}$ & $\alpha$ & $\delta$ & Parameters & TOAs \\
             number      & (Hz)  & ($10^{-15}$ Hz s$^{-1}$) & (hh:mm:ss.s) & ($^\circ:\arcmin:\arcsec$)  & fitted & fitted \\
		\hline
		1 & 541.8936548(11) & 0*        & 00:24 05.67* & $-$72:04:52.62* & $\nu$ & 1 group\\
        2 & 541.89365529(39)  & 0*        & 00:24:05.67* & $-$72:04:52.62* & $\nu$ & Initial groups + JUMPS \\
		3 & 541.89365651(5)  & 0*        & 00:24:07.1(8) & $-$72:05:17.0(36) & $\nu, \alpha, \delta$ & Initial groups + JUMPS\\
		4 & 541.8936564(6)  & 12(6)    & 00:24:07.0(8) & $-$72:05:19.9(39) & $\nu, \alpha, \delta, \dot{\nu}$ & Initial groups + JUMPS \\
        5 & 541.8936568(6)  & 4(4)    & 00:24:07.4(8) & $-$72:05:18.6(35) & $\nu, \alpha, \delta, \dot{\nu}$ & 1-day groups + JUMPS \\
        6 & 541.89365494899(18) & 13.475(4) & 00:24:07.2783(10) & $-$72:05:19.521(4)& $\nu, \alpha, \delta, \dot{\nu}, \mu_\alpha, \mu_\delta$ & All connected \\
		\hline
	\end{tabular}
\end{table*}

\subsection{Rotation count between observations}

Although the rotation count within an observation is, as we have seen,
generally well known, this is not the case for the rotation count
between more widely spaced TOAs.

We can see this by repeating the procedure in the previous section for
all available observations:
\begin{enumerate}
\item Make TOAs for all observations where the pulsar has appeared.
Whenever possible, make at least 3 TOAs - this is important for
obtaining residual slopes from each day's data.
\item Use the same ephemeris as in the previous step,
\item Run {\sc tempo} and look at the residuals.
\end{enumerate}

For the ephemeris listed above, the pre-fit residuals are displayed in
Fig.~\ref{fig:initial_residuals}. These are scattered evenly through the
whole spin phase, from $-0.5$ to $0.5$, with no discernible pattern.
This means that our first
ephemeris is not precise enough to predict the rotation count
between all these TOAs, otherwise all the residuals
would be confined to a narrow range of phases,
as in Fig.~\ref{fig:first_residuals}. Thus, the starting ephemeris
is a {\em partial} ephemeris: it is certainly not phase-coherent.

This lack of precision stems from a variety of factors:
a) Lack of precision of the spin frequency $\nu$ (this is precise
enough to predict TOAs for a few hours, but probably not enough 
to do so for many years) b) the possibility that
the sky position of the pulsar is slightly different from the
position of the centre of 47~Tuc; this introduces
small delays due to the Earth's motion that are nonetheless
larger than one rotation of the pulsar and c) the unknown variation of
$\nu$ with time. In what follows, we will assume at first that
this can be described by a single spin frequency derivative, $\dot{\nu}$.

In both plots of Fig.~\ref{fig:initial_residuals}, we see that whenever
multiple TOAs are derived from successive sub-integrations
within the same observation, their residuals appear in (nearly) horizontal groups,
except for two cases that appear near phase 0.5 that will be discussed below.
This means that those groups of TOAs are connected
(see Section~\ref{sec:phase_within_observation}).
The time interval between two such groups is a {\em gap}; thus by definition
the rotation count within a gap is not necessarily
known at this stage.

\subsection{Moving forward: a better ephemeris}
\label{sec:moving_forward}

The first step toward the determination of the rotation count
is to make a more precise estimate of the spin frequency
and other parameters using {\em all} connected TOA groups, right from the start.
This has the great advantage that, for every step of the work, the estimates of
all timing parameters are constrained by the residuals slopes from all
connected TOA groups.

Since the initial ephemeris does not provide a reliable rotation count across all
gaps, we cannot assume to know it. We can remove this assumption in {\sc tempo}
by fitting an arbitrary time offset for each group of connected TOAs.
This can be achieved -- in {\sc tempo} -- by bracketing the TOAs from all observations
but one (in this case the last) with JUMP statements, as exemplified here:

\begin{scriptsize}
\begin{verbatim}
MODE 1

JUMP
7               1390.000 51413.6357638815603   18.173                 0.00000
7               1390.000 51413.6499999905185   23.038                 0.00000
JUMP

JUMP
7               1390.000 51413.6895833852018   14.824                 0.00000
7               1390.000 51413.7381944448891   14.797                 0.00000
7               1390.000 51413.7868055551721   18.587                 0.00000
7               1390.000 51413.8354166713418   22.973                 0.00000
JUMP

JUMP
7               1390.000 51490.5017355131067   24.352                 0.00000
JUMP

...

(some TOAs not shown here)

...
C DON'T BRACKET WITH JUMPS
7               1390.000 54816.4975696298174   31.740                 0.00000
\end{verbatim}
\end{scriptsize}

\begin{figure}
	\includegraphics[width=\columnwidth]{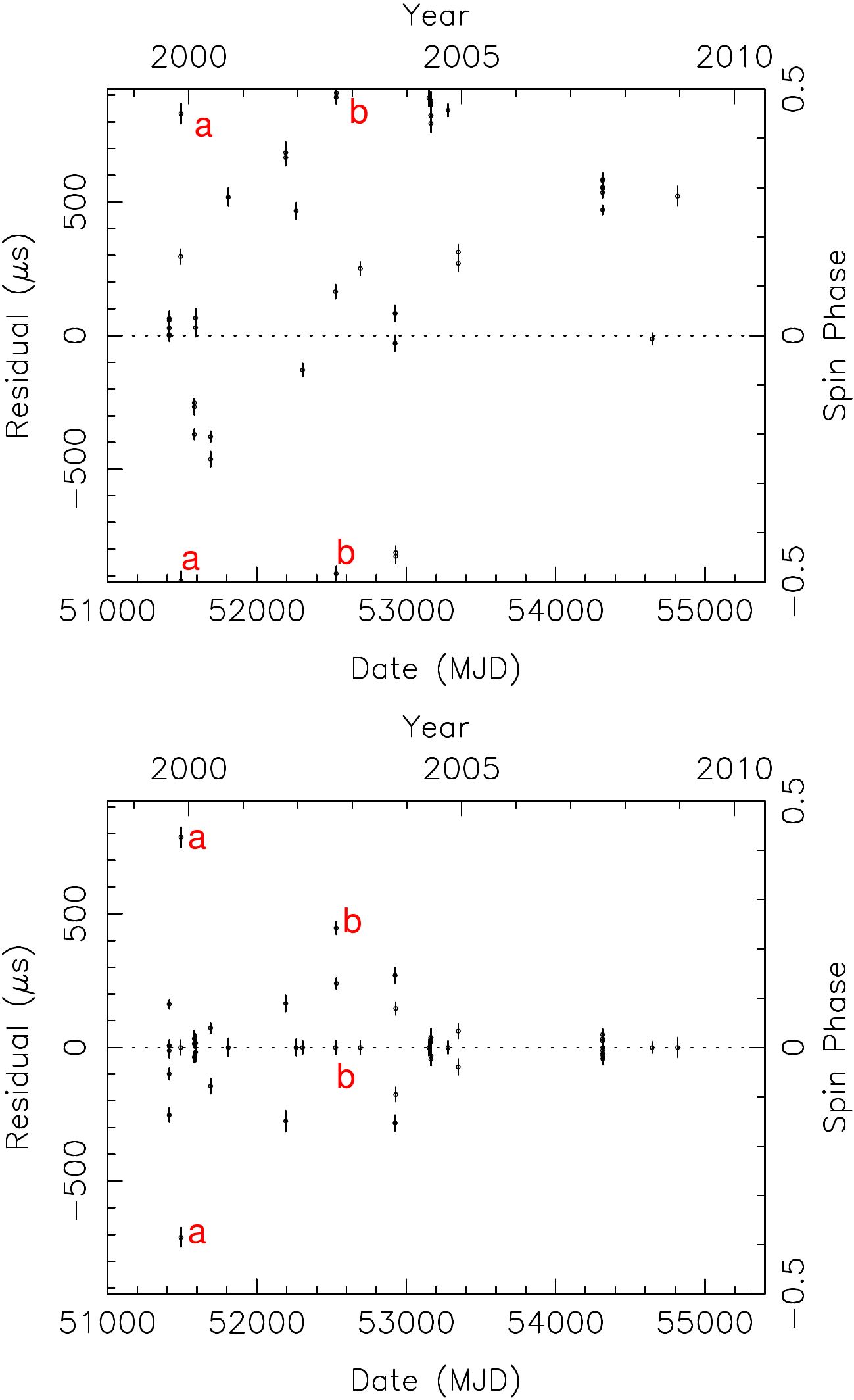}
    \caption{Pre-fit (top) and post-fit (bottom) 
    residuals for all TOAs for 47~Tuc~aa as a function of epoch.
    The pre-fit residuals are scattered evenly across the full
    spin phase space (between $-0.5$ and $0.5$), i.e., our starting ephemeris
    is not precise enough to predict the phase even approximately. However, for TOAs
    derived from a single observation (i.e., those where we know there is connection),
    the residuals appear in nearly horizontal groups, this means that
    the predicted spin period for each observation is accurate.
    If a group of residuals appears near phase 0.5, then some residuals in that group
    might wrap around and appear near phase $-$0.5. So both TOAs indicated with ``a''
    and both indicated with ``b'' really appear very close to each other in phase.
    Because of the mis-identification of the rotation number, the post-fit reduced $\chi^2$
    is 114.9.
    }
    \label{fig:initial_residuals}
\end{figure}

Then, as before, run {\sc tempo}. The pre-fit residuals are the same as
displayed in the top of Fig~\ref{fig:initial_residuals}.

Running {\sc tempo} at this stage, we obtain the residuals
displayed at the bottom of Fig.~\ref{fig:initial_residuals}. These have
a very high reduced $\chi^2$ of 225.15.
The reason for this (and this is something we should always beware
during this process) is that there are groups of residuals close to the spin phase
of 0.5: individual residuals in such groups might appear at a spin phases
near 0.5 {\em and} $-0.5$. This is happening for the
two pairs of TOAs called ``a'' (at MJD = 51492) and ``b'' (at MJD = 52531)
respectively. In this situation, the wrong rotation count
is being assumed within those observations, so {\sc tempo} cannot produce
a good estimate of the parameters. In order to correct this, one should
introduce or subtract (whichever is suitable)
an extra rotation in between the TOAs in each group,
so that they appear with the same phase in the
pre-fit residual plot. This can be done with a PHASE +1/$-$1 statement
in the TOA list, for the cases mentioned below these are:

\begin{scriptsize}
\begin{verbatim}
(...)
JUMP
7               1390.000 51492.5850727545785   31.229                 0.00000
PHASE -1
7               1390.000 51492.6440955080127   32.860                 0.00000
JUMP
C Optional
PHASE +1
(...)
\end{verbatim}
\end{scriptsize}
and
\begin{scriptsize}
\begin{verbatim}
(...)
JUMP
7               1390.000 52531.6708365991755   24.719                 0.00000
PHASE -1
7               1390.000 52531.7347222220090   20.900                 0.00000
7               1390.000 52531.7968733372661   17.472                 0.00000
JUMP
C Optional
PHASE +1
(...)
\end{verbatim}
\end{scriptsize}

\begin{figure}
	\includegraphics[width=\columnwidth]{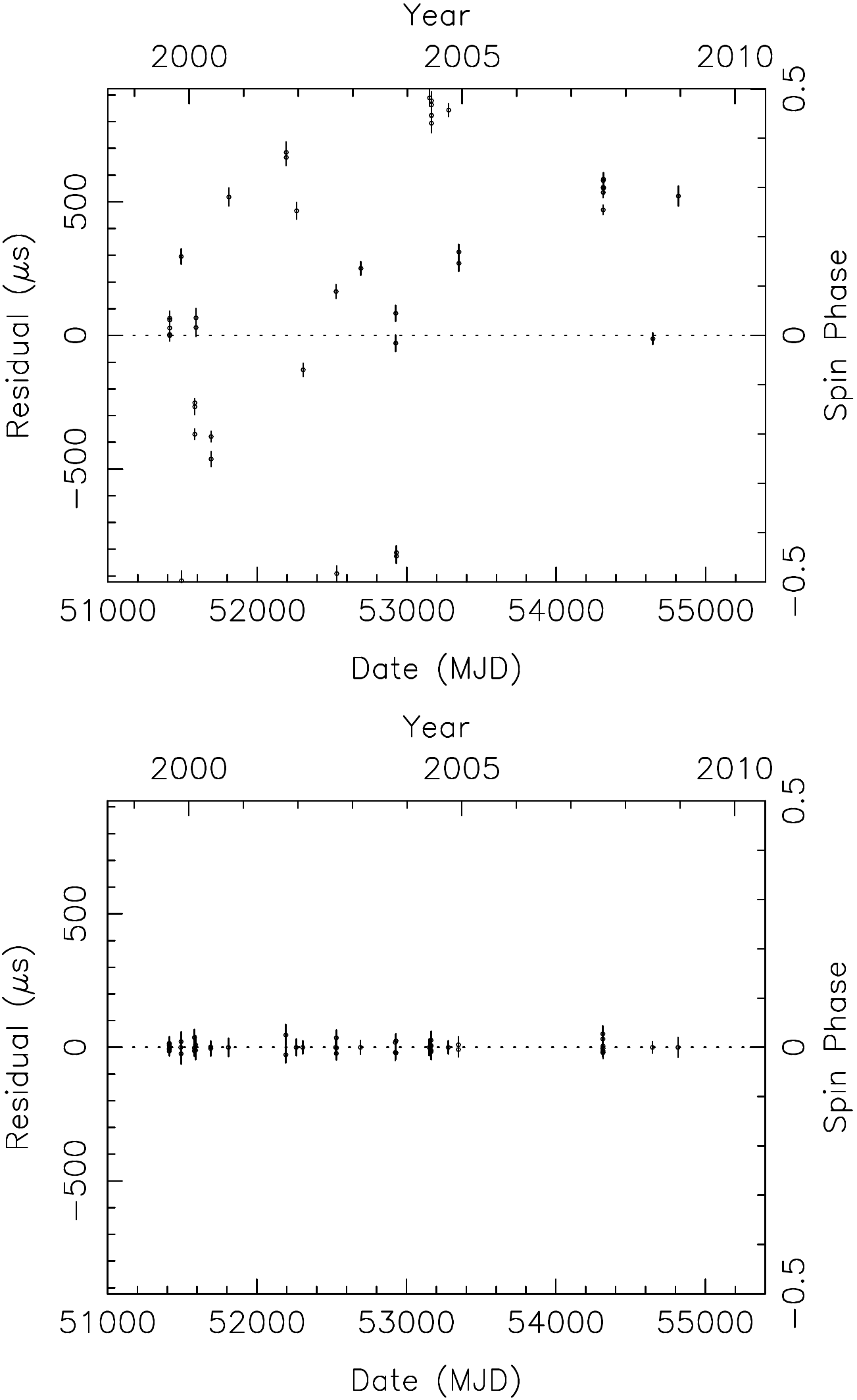}
    \caption{Pre-fit  and post-fit residuals for all TOAs for 47~Tuc~aa as a function of
    epoch, now with the correct rotation count for epochs ``a'' and ``b''.
    Although the pre-fit residuals are scattered evenly across the full
    spin phase space (between $-0.5$ and $0.5$), the post-fit residuals
    all appear near phase zero, with a reduced $\chi^2$ near 1.9.
    This is partially a consequence of the fact that we
    are fitting an arbitrary time offset to each observation. Flattening
    the residuals slopes within individual observations constrains the timing
    parameters.
    }
    \label{fig:initial_JUMPs}
\end{figure}

The optional PHASE +1 statements were introduced in both cases so that all
residual phases continued being displayed in the interval from $-$0.5 to 0.5.
Only after this correction, i.e., only after finding the correct rotation count
within those observations, can {\sc tempo} converge on an ephemeris with a
reduced $\chi^2$ close to 1.
The pre and post-fit residuals as a function of epoch obtained
after this correction are presented in Fig.~\ref{fig:initial_JUMPs}.

In the lower plot of Fig.~\ref{fig:initial_JUMPs}, all post-fit residuals appear within a
narrow window of spin phases.
This is mostly (but not entirely) a consequence of the fact that we have fitted time offsets
for each group of connected TOAs. Solely removing these offsets does not eliminate the
residual slopes for each group of connected TOAs; this is done by fitting
the timing parameters. 

Let us now exemplify this. If we fit only for $\nu$, we obtain the spin frequency
(presented in the second row in Table~\ref{tab:parameters}) that is more than twice as
precise as the spin period estimate obtained from the best observation in
Section~\ref{sec:phase_within_observation}.
However, this ephemeris produced residuals with a reduced $\chi^2$ of 3.79 -- still
with non-zero residual slopes within each day -- so we might
need to fit for (an)other parameter(s).

Because the total time span is larger than one year, it is possible to fit for $\nu$, $\alpha$ and $\delta$, as the covariance between these parameters can be broken. If we do so, we obtain residuals
with a reduced $\chi^2$ of 1.54,
which is a major improvement. The solution we obtain then is presented
in the third row of Table~\ref{tab:parameters}. We can see here that,
although the spin frequency is not quite as precise as in the previous column,
we can already pinpoint the location of the pulsar with a precision of
a few arcseconds, which is a major improvement over the previous
(and, as we can now see, correct) assumption
that the pulsar is near the centre of 47~Tuc.

If we fit for $\nu$, $\alpha$ and $\delta$ {\em and} $\dot{\nu}$, then
the reduced $\chi^2$ decreases to 1.40. The resulting best ephemeris
is presented in the fourth row of Table~\ref{tab:parameters}.
Although the resultant $\dot{P} \equiv - \dot{\nu} / \nu^2$ is not very significant
$(-4.0 \, \pm \, 2.0)\, \times \, 10^{-20}\, \rm s \, s^{-1}$, it is well within the
range of what one finds for other pulsars in 47~Tuc.
This implies that this parameter should be fit at this stage.
Our starting ephemeris file is then the same as in Section~\ref{sec:phase_within_observation},
only with a few extra fitting flags (1's) added after $\alpha$, $\delta$ and $\dot{\nu}$ lines.

To summarize: although we have not determined the rotation count for
a single gap, by using the JUMP statements we can use the
information contained in the residual slopes within all connected TOA
groups to derive a reasonably precise position, period, and even a
strong constraint on $\dot{\nu}$. These estimates cannot be derived from any individual group of TOAs.

\subsection{Connecting the gaps}
\label{sec:testing}

We now arrive at the most important step on the path to get the
correct global rotation count for the pulsar. We look at the TOA
list and find closely spaced groups of connected TOAs, i.e.,
short gaps. As an example, for the best detection of 47~Tuc~aa,
there is an observation earlier that
day where the pulsar was also detected. Here are the TOAs:

\begin{scriptsize}
\begin{verbatim}
(...)

JUMP
7               1390.000 54314.7031249750801   15.335                 0.00000
7               1390.000 54314.7515008653730   25.434                 0.00000
JUMP

JUMP
7               1390.000 54314.7954878129250   16.660                 0.00000
7               1390.000 54314.8211805391308   17.735                 0.00000
7               1390.000 54314.8468749751720   15.921                 0.00000
7               1390.000 54314.8725694346731   21.125                 0.00000
7               1390.000 54314.8982638745009   19.083                 0.00000
7               1390.000 54314.9222205617143   19.950                 0.00000
JUMP

(...)
\end{verbatim}
\end{scriptsize}
The question is now: is our original ephemeris (presented in
Section~\ref{sec:phase_within_observation}) precise enough to predict the
rotation count for this gap or not?

To find out, we comment out the inner pair of JUMP statements and introduce
a PHASE +N statement, where N is an integer that determines
the correction to the rotation count predicted by the ephemeris.
In what follows, such integers (or combinations of $k$ integers
when discussing $k$ gaps) are referred to as {\em $k$-gap solutions},
a 1-gap solution will be referred to simply as a {\em solution}.

\begin{scriptsize}
\begin{verbatim}
(...)

JUMP
7               1390.000 54314.7031249750801   15.335                 0.00000
7               1390.000 54314.7515008653730   25.434                 0.00000
C JUMP
PHASE +0
C JUMP
7               1390.000 54314.7954878129250   16.660                 0.00000
7               1390.000 54314.8211805391308   17.735                 0.00000
7               1390.000 54314.8468749751720   15.921                 0.00000
7               1390.000 54314.8725694346731   21.125                 0.00000
7               1390.000 54314.8982638745009   19.083                 0.00000
7               1390.000 54314.9222205617143   19.950                 0.00000
JUMP

(...)
\end{verbatim}
\end{scriptsize}

\begin{figure}
	\includegraphics[width=\columnwidth]{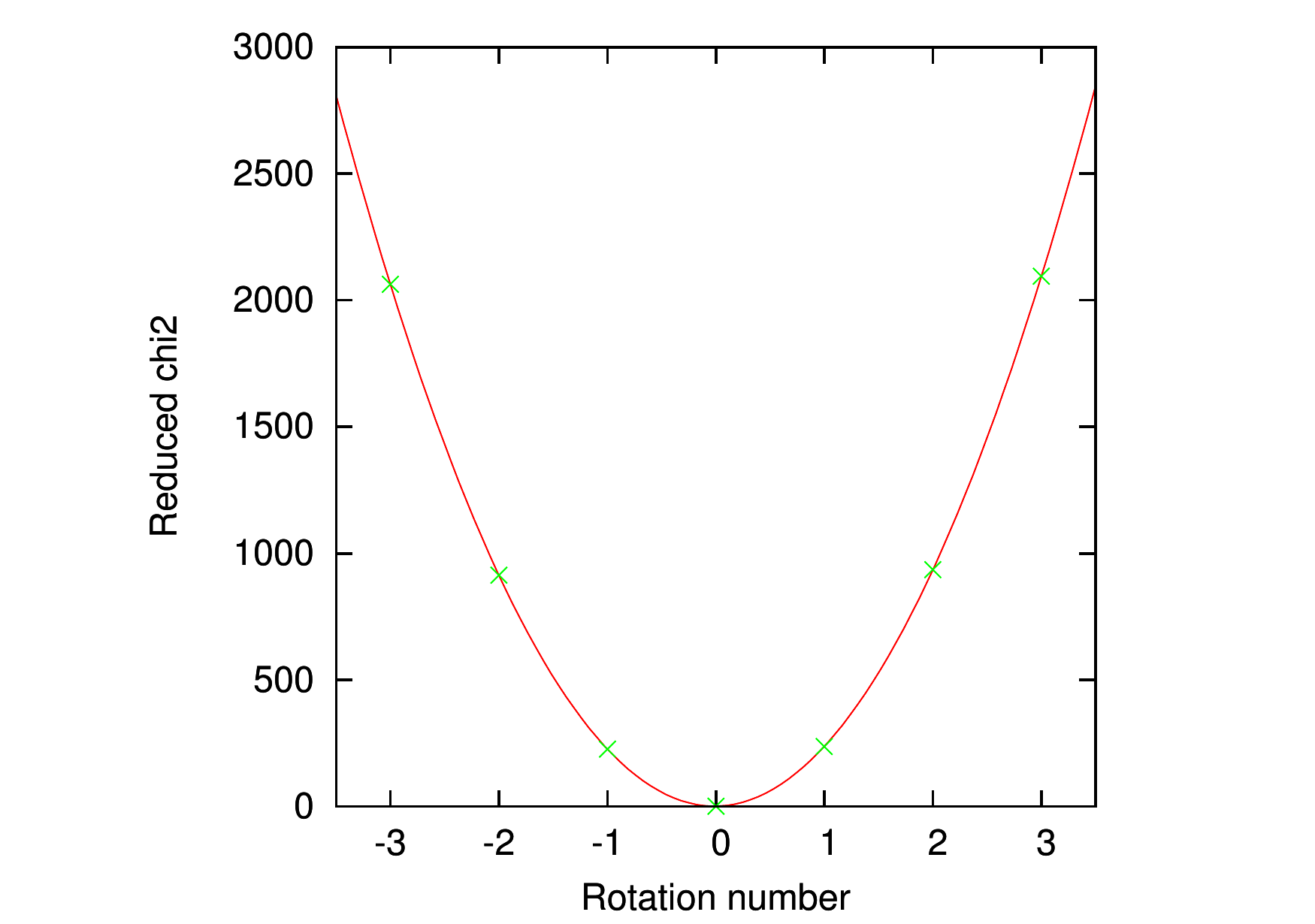}
    \caption{Reduced $\chi^2$ as a function of the solution for the
    small gap in MJD = 54314.
    The values derived by {\sc tempo} (green crosses) follow closely a 
    parabolic curve; the latter is represented by the solid red line.
    }
    \label{fig:parabola}
\end{figure}

Now we run {\sc tempo}, still with the original ephemeris.
Assigning the integers $-3$, $-2$, $-1$, 0, 1, 2 and 3 to N,
we obtain the following values for the reduced $\chi^2$:
2063.0, 914.1, 226.9, 1.37, 237.5, 935.3 and 2094.8
(see Fig.~\ref{fig:parabola}).

The quality of the fit is much
higher for N = 0 (with a reduced $\chi^2$ of 1.37) than for the second
best fit ($-1$, where the reduced $\chi^2 = 226.9$). In such cases we can say
with great confidence that there is a unique {\em acceptable solution}, ``0''.
This means that the rotation count of
this gap is {\em unambiguous}, and in this case identical
to the rotation count predicted by the initial ephemeris.

In this situation we can eliminate the gap, i.e.,
declare the groups on both sides to be mutually
connected (in what follows, we simply describe this
as {\em connecting the gap}). Because the gap has been connected,
there is no need to reintroduce its JUMP statements in TOA list, but
whatever PHASE +N statement worked best should stay.
This connection results in more tightly constrained timing parameters.

In Fig.~\ref{fig:parabola}, we can see that the relation between the
reduced $\chi^2$ and $N$ follows very closely a parabola, which is depicted by
the solid red line. In the sections that follow, we make use of this fact.

\subsection{Connecting the remaining gaps}
\label{sec:full_data_set}

After this stage, we look for other gaps shorter than 1 day.
The two gaps we found (within MJD 51413 and 53164) also
have unique solutions, in both cases 0, so both gaps can be connected.
Since we now have slightly larger connected TOA groups, the post-fit
timing parameters will be more precise. However,
as we can see in the fifth row of Table~\ref{tab:parameters}, the
improvement in this case is marginal.

In order to proceed, we must try to find other gaps with unique solutions,
since such gaps can be immediately connected. The longer connected dataset
will then result in an ephemeris with more tightly constrained
parameters; the latter will more likely make other gaps
unambiguous. This is a runaway process, which means that,
using this technique, we can, for most pulsars, determine unique solutions
for all gaps, i.e., determine their global rotation count.
This can be done even if the detections appear to be very sparse at first -- all that matters
is that there is a good distribution of {\em gap lengths}.

We will mention here a technical detail of this process, which will
become apparent to anyone using this technique.
Although the solutions for the shorter gaps
will be relatively small integers, as the length of the gaps
increases, the magnitude of the solutions increases significantly. In order to
mitigate the problem, the initial ephemeris can be replaced with the
post-fit ephemeris calculated by {\sc tempo} at any stage. 
For instance, we could have replaced the initial guess ephemeris
in Section~\ref{sec:phase_within_observation} with the better
ephemeris derived in the previous section. For this to work,
all previous PHASE +N statements in the TOA list must
be commented out, since the corresponding rotation count is already
taken into account by the new starting ephemeris. After such a change, however,
we must always be careful about groups of TOAs at phases close to 0.5,
as mentioned in Section~\ref{sec:moving_forward}.

Another important point to keep in mind is that for this
technique to succeed, three conditions must be met:

a) The TOAs must be accurate. If one TOA is for some reason in error, it will
derail the whole process. For this reason, we must be careful with the
selection of the pulse profiles while deriving TOAs.

b) The TOAs should be consistent with each other (i.e., preferably come from
the same telescope and timing instrument), also the template used to derive them
should be the same and

c) the pre-fit
ephemeris must contain an appropriate set of timing parameters.
For many isolated pulsars, $\alpha$, $\delta$, $\nu$ and $\dot{\nu}$
are perfectly adequate; as we will see this is also the case for 47~Tuc~aa.
For longer timescales, proper motion parameters might be necessary.
However, even among isolated pulsars, particularly the younger ones,
there are phenomena like glitches
and timing noise that can greatly complicate the analysis.

When sparsely detected pulsars are part of binary systems (something that becomes
immediately apparent from the variation of the spin period), we can make a
preliminary determination of the orbital parameters from the observed spin periods
and their derivatives using the method presented by \cite{2001MNRAS.322..885F}
\footnote{The related {\sc CIRCORBIT} software can be found in
\url{http://www3.mpifr-bonn.mpg.de/staff/pfreire/programs/circorbit.tar}
};
this was used to find the orbits of several pulsars in 47~Tuc.
Those initial orbital parameters can then be refined by fitting the observed
Doppler-shifted spin period as a function of time to a Keplerian model
(something that can be done using a program like {\sc FITORBIT}).

Phase-coherent ephemerides for such pulsars can be derived using the method
we have just described above; however, to achieve that we must exercise good judgment
on the binary parameters that must be fit. For a pulsar in wide, nearly circular orbit
with a compact object, it is likely that fitting for Keplerian parameters
from the start will be enough to find phase connection. If the orbit is compact
and eccentric, then fitting for the rate of advance of periastron
($\dot{\omega}$) from the start might be necessary. In the case of
eclipsing binary pulsars, then unpredictable changes in the orbital period
must also be taken into account. 

{ Back to 47~Tuc~aa, and before we continue, we note that at this stage of
the work the reduced $\chi^2$ of the residuals is 1.40. Our previous
experience timing other faint MSPs suggests this is likely to be caused by
a slight under-estimate of the TOA uncertainties.
We can fix this by re-scaling the latter.
In {\sc tempo}, this is done by introducing a flag in the TOA list called
EFAC, followed by a numerical value. All TOA uncertainties are then
multiplied by this factor. In what follows, we use an EFAC of
$\sqrt{\chi^2_{\rm red}} \, \simeq \,1.185$. In this way
the reduced $\chi^2$ becomes 1.0. This (slightly) increases the uncertainties
of the timing parameters to more realistic values.}

\section{Sieving solutions}
\label{sec:mutiple_trials}


Apart from the single-day
gaps mentioned in Section~\ref{sec:full_data_set}, there are, in the case
of 47~Tuc~aa, no wider gaps with unique solutions, i.e., all remaining gaps
are {\em ambiguous}. In such cases it is not clear how to proceed.

In what follows, we will address this problem.
We start by presenting an automatic method for finding and listing
all acceptable solutions for each gap, this is called the {\em gap mapper}.
By acceptable solutions we mean a solution having a
reduced $\chi^2 < 2.0$, where that factor should not
be too large as to allow an exceedingly large number of
solutions, but it should not be so small as to exclude the
correct solution, which might have a reduced $\chi^2$ larger than
1.0. As we will see in Section~\ref{sec:finding}, the eventual
correct global solution has, using the four parameters being fit,
a reduced $\chi^2\, = \, 1.39$ (already using the EFAC factor of 1.185
mentioned above). Using a smaller $\chi^2$ threshold
might have excluded the correct solution. The limit of 2.0 is chosen
because for most MSPs the reduced $\chi^2$ in their published timing
solutions is smaller than that.

\subsection{Mapping a gap}
\label{sec:mapping}
One tool we will need to do the connection is to learn
how to find the best solutions for any gap automatically.
An efficient method to achieve this can be derived from Fig.~\ref{fig:parabola}
in Section \ref{sec:full_data_set}: the reduced $\chi^2$ varies as a parabola.
We can find the minimum of a parabola if we sample it
(i.e., use {\sc tempo} to evaluate the reduced $\chi^2$) for three
(not necessarily acceptable) solutions.
We generally choose these to be $-b$, 0 and $b$, with $b = 5$ being a
good choice. In this case, then the minimum will be the integer $M$ closest to
\begin{equation}
\label{eq:minimum}
m = \frac{b}{2} \frac{\chi^2(-b) - \chi^2(b)}{\chi^2(b) + \chi^2(-b) - 2 \chi^2(0)}.
\end{equation}
Then, starting from $M$, we evaluate the integers immediately above and below
it to establish the quality of fits, keeping a record of the acceptable
solutions.

Using this simple device, we have mapped
all the gaps in our 47~Tuc~aa data set. For each gap $i$ we list in Table~\ref{tab:gaps}
the best solution ($M_i$) and the number of acceptable 1-gap solutions ($n_i$) around it
(roughly one half of that number of solutions above and below $M_i$).
\begin{table}
	\centering
	\caption{Unconnected gaps in the 47~Tuc~aa TOA data set 
    and corresponding best solution for each gap ($M_i$) and number of acceptable
    solutions ($n_i$).\label{tab:gaps}}
	\begin{tabular}{r c r r} 
		\hline
		$i$ & Gap & $M_i$ & $n_i$\\               
		\hline
		1 & 51413 - 51490 & $-$14 & 58 \\
        2 & 51490 - 51492 & 0 & 2 \\
		3 & 51492 - 51582 & $-$53 & 86 \\
		4 & 51582 - 51589 & $-3$ & 7 \\
        5 & 51589 - 51690 & +8 & 115 \\
        6 & 51690 - 51809 & +34 & 84 \\
        7 & 51809 - 52193 & $-$38 & 192 \\
        8 & 52193 - 52262 & $-38$ & 72 \\
        9 & 52262 - 52307 & $-$27 & 46 \\
        10 & 52307 - 52526 & +33 & 168 \\
        11 & 52526 - 52531 & 0 & 3 \\
        12 & 52531 - 52693 & $-$81 & 146 \\
        13 & 52693 - 52925 & +33 & 140 \\
        14 & 52925 - 52930 & $-1$ & 5 \\
        15 & 52930 - 53152 & $-$70 & 172 \\
        16 & 53152 - 53164 & +5 & 13 \\
        17 & 53164 - 53280 & +22 & 49 \\
        18 & 53280 - 53348 & $-$36 & 80 \\
        19 & 53348 - 54314 & $-$128 & 697 \\
        20 & 54314 - 54646 & $-$74 & 304 \\
        21 & 54646 - 54816 & $-$37 & 168 \\
		\hline
	\end{tabular}
\end{table}

\subsection{Sieving}
\label{sec:sieving}

{ One possible way of finding the timing solution for 47~Tuc~aa would be to
try all possible combinations of acceptable solutions for the 21 gaps.
However, this would clearly be impractical:
the product of all $n_i$ for this data set is
\begin{equation}
6\,  894\, 108\,  365\, 755\, 450\, 688\, 986\, 274\, 510\, 929\, 920\, 000, \nonumber
\end{equation}
or approximately $6.9 \, \times \, 10^{36}$}. While mapping these
gaps each {\sc tempo} iteration took
about 0.17 seconds; this means that, with the same computer, the whole process would take 
$3.7 \, \times \, 10^{28}$ years. As we will see below, the vast majority of these
calculations are unnecessary.

We now describe the fundamental concept of the solution-finding algorithm. Although
in the previous step we are not able to find a single unambiguous
gap, we can assume, for one of the gaps (say, number 2), that one of
its two acceptable 1-gap solutions ($-$1 and 0) is in turn correct. For each of these
assumptions we try to find (using the gap mapper described in the previous section)
acceptable solutions for a second gap, let's say number 11
(this gap should be chosen in order to yield the smallest possible number of
acceptable solutions, i.e., it should be a short gap).
If we find such a solution, then together with the solution for gap 2 being assumed this
makes a new 2-gap solution. Then, assuming each 2-gap solution we proceed to
search for acceptable solutions for the next gap (say, number 4), thus finding
3-gap solutions. We continue adding more gaps until one single $k$-gap
solution stands out in terms of its quality.

The reason why this works is because of what has been described in
Section~\ref{sec:connection}: the assumption that there is phase connection for the
first gap (in this case number 2) increases the precision of the ephemeris,
which is derived from a larger group of connected TOAs.
This means that, when we attempt to connect a new gap (in this case 11),
the number of acceptable solutions decreases (in this case from three to {\em two} for
each of the 2 assumptions).

This means that there are acceptable 1-gap solutions for
gaps 2 and 11 that do not provide an acceptable solution when they are assumed for
both gaps at the same time, i.e., they don't work well together as a 2-gap solution.
Since the reduced $\chi^2$ of this 2-gap solution is high, it is unlikely that any
$k$-gap (with $k \, > \, 2$) solutions based on this 2-gap solution will have a low
reduced $\chi^2$. We can therefore forget about this 2-gap solution when attempting
to find the rotation count for the following gaps, i.e., the combination does not pass
the $\chi^2$ sieve.

As we will see in Section~\ref{sec:finding} for the specific case of 47~Tuc~aa,
this sieve is extremely powerful because it decimates the number of possible
solutions every time we test the surviving $k$-gap solutions 
against gap $k + 1$. The effect might appear to be small at every iteration, but it increases
exponentially with every stage.
The number of acceptable solutions increases fast at first,
but as more gaps are tested, the number of acceptable
solutions stabilizes and then starts decreasing. Eventually one solution
starts standing out in terms of the quality of the fit.
This solution is then generally able to unambiguously connect all other gaps.

\subsection{Implementation and future improvements}
\label{sec:parallel}
The relative simplicity of the method outlined above allows a simple implementation as a
UNIX shell script (called sieve.sh); this is freely available in the Dracula Github
repository\footnote{\url{https://github.com/pfreire163/Dracula}}, complete with a short
description of its usage. The advantage of such
a script is that it can run on any UNIX or LINUX platform (which normally have {\sc sed} and
{\sc awk} by default), without the need for any special
program other than {\sc tempo}.

However, there are still many improvements that could be made to speed up the process
and make it more robust.

{ One improvement that has already been partly implemented is that the script can be made
to run much faster by running in shared memory, as in the case
of our Github implementation. This avoids writing output files to hard disk, a process that
causes disk wear and takes significantly more time than the calculations themselves.}

Two further improvements have to do with limitations of {\sc tempo}.
The first one is that the values of the reduced $\chi^2$ being reported have a limited range:
if they are larger than 999999.99 they are written by the program as a string of asterisks.
In such cases it is impossible to determine the position of the minimum using
eq.~\ref{eq:minimum}. Second, each time {\sc tempo} is called
it spends most of the time a) reading clock correction files and correcting
the TOAs, b) reading Earth rotation files and
calculating the observatory position and c) reading Solar System ephemerides and calculating
the Earth position. These operations only need to be done once.
{ We note that the {\sc PINT} timing program does not have these limitations,
and for that reason it will certainly allow much faster discovery of rotation counts.

A fourth improvement has to do with the script itself. In the current version
the user decides which gap is to be connected next. This is not necessarily
the gap that yields the smallest number of solutions. An obvious next step is
to have the script determine what is the best gap to connect next.
This minimization is likely to be very important,
since any reduction in the number of solutions at each step has
exponential consequences when many steps are considered.

A fifth improvement has to do with an analysis of the timing solutions
that are produced with every fit -- the information they provide has
not been used in the work above. If the estimated positions are
well outside the telescope beam, or if the pulsar has a negative
$\dot{P}$, then the solutions can be safely excluded.
However, the latter condition must be used with care:
in a globular cluster, many pulsars have negative $\dot{P}$;
as we shall see this is the case for 47~Tuc~aa.
}

Another technique that should provide an extreme acceleration is to parallelize the
computation of solutions for successive gaps. This is conceptually very simple:
while a version of the script is still searching for solutions
for gap A (using the gap mapper), it passes any acceptable 1-gap solutions it finds
to a second version of the script working on gap B.
This sorts (according to reduced $\chi^2$) the new 1-gap solution against all previous
unprocessed 1-gap solutions that have been passed to it from gap A
and then works on the best one (deleting it from the list of tasks to do).
Assuming that 1-gap solution, it will use the gap mapper to look for
solutions for gap B. If it finds any acceptable 2-gap solutions,
it will pass them to an analogous script using the gap mapper to find
solutions for gap C. This script passes the acceptable 3-gap
solutions it finds to an analogous script working on gap D, etc.
This saves a lot of time not only because of the parallelism
- there is no need to wait for the conclusion of the discovery of all
acceptable $(k-1)$-gap solutions to start searching for $k$-gap solutions
- but also because the
sorting by reduced $\chi^2$ at every stage focuses the processing on the best
possible solutions first at every stage. This should lead to a
bare minimum of {\sc tempo} iterations.

\section{Timing solution of 47~Tuc~aa}
\label{sec:47Tucaa}

\begin{table}
	\centering
	\caption{Successive gaps (order given by $k$)
    being tested for 47~Tuc~aa; see $i$ numbers in Table~\ref{tab:gaps}.
    $P$ - product of the $n_i$ for the gaps
    being tested, which is the number of solutions we would have to test
    if we were not sieving the acceptable solutions at each
    stage. $N_k$ - number of $k$-gap solutions actually being tested at each stage.
    Here we can see how this value becomes progressively
    smaller than $P$, demonstrating the power of the sieve technique.
    In the last column, we indicate the correct solution for these gaps
    (number 2 of the 30 solutions at step 10).
    These are small corrections to the rotation count predicted by our initial
    ephemeris.
    \label{tab:successive}}
	\begin{tabular}{r r l l r r} 
		\hline
		$k$ & $i$ & $n_i$ & $P$ & $N_k$ & Solution \\
        \hline
        1 & 2  & 2 & 2 & 2 & 0 \\        
        2 & 11 & $\times$ 3 & = 6 & 4 & 1\\        
		3 & 4  & $\times$ 7 & = 42 & 16 & $-$2 \\
        4 & 14 & $\times$ 5 & = 210 & 22 & $-$1\\        
        5 & 16 & $\times$ 13 & = 2730 & 26 & 6 \\
        6 & 17 & $\times$ 49 & = 133770 & 36 & 31 \\       
        7 & 18 & $\times$ 80 & = 10701600 & 90 & $-$33\\        
        8 & 1  & $\times$ 58 & = 620692800 & 135 & 0 \\      
  		9 & 3  & $\times$ 86 & = 53379580800 & 84 & $-$39 \\
        10 & 5  & $\times$ 115 & = 6138651792000 & 32 & 25 \\
		\hline
	\end{tabular}
\end{table}

Using the technique described in Sections~\ref{sec:mapping} and \ref{sec:sieving}, we were
able to determine a phase-coherent ephemeris for 47~Tuc~aa with 1278 {\sc tempo}
iterations used in the gap mapper, plus an extra 476 {\sc tempo} iterations
that evaluated the quality of valid solutions around minima, plus a similar
number of iterations that evaluated non-valid combinations. The first two sets of iterations
took a total time of about 3 minutes. In what follows, we give a detailed account of
this process and describe the resulting phase-coherent ephemeris.

\subsection{Finding the rotation count for 47 Tuc aa}
\label{sec:finding}

First, we ran the sieve.sh script looking for solutions for a single gap
(gap n.2 in Table~\ref{tab:gaps}), using the gap mapper.
There are only two 1-gap solutions, ($-1$) and (0). 
Then we ran the same script again to find solutions
for gap 11 based on the two 1-gaps solutions for gap 2.
It found four acceptable 2-gap solutions: (0, 1), ($-1$, 0),
($-1$, $-1$) and (0, 0). Then, for each of these the script
searches for acceptable
solutions for gap 4, finding 15 acceptable 3-gap solutions.
The successive gaps being tested and the number $N_k$ of acceptable
$k$-gap solutions found are listed in Table~\ref{tab:successive}.

The number of $k$-gap solutions the script has to test at each stage becomes progressively
insignificant compared to $P$, which is the number of all possible
combinations of integers we would have to test
if we were not sieving the acceptable solutions at each stage.
This number, $P$, is the product of the number of 1-gap solutions ($n_i$) for the
gaps already considered. This is a clear demonstration of the power of sieving.

After finding solutions for the tenth gap
(number 5 in Table~\ref{tab:successive}), the script
starts running into a problem mentioned in Section~\ref{sec:parallel}: the values
of reduced $\chi^2$ at one or more of the sample
points used by the gap mapper ($-b$, 0 and $b$) start becoming too
large. This has a good implication:
it means that the ephemerides resulting from the assumption of
these 10-gap solutions have a strong predictive power, otherwise the
values of reduced $\chi^2$ for ``wrong'' solutions would not be so large.
This is the stage at which we should start looking in more detail at the individual
solutions. Fortunately at this stage there are only 32 acceptable 10-gap solutions to look at.

We wrote a second script (test.sh, also found in the Dracula Github repository)
that, for any set of $k$-gap solutions, applies the PHASE statements contained in each of them
and then uses {\sc tempo} to derive an ephemeris that assumes that rotation count.
Then the script switches to the latter
as a starting ephemeris, for this to work it removes all the PHASE
statements from the previous TOA list since those are already taken into account by the new
ephemeris.
If all pre-fit residuals obtained with this ephemeris fall
within a relatively narrow phase range, or if they have a clear, slowly-varying
long-term pattern, then we know we have the correct solution.

\begin{figure}
	\includegraphics[width=\columnwidth]{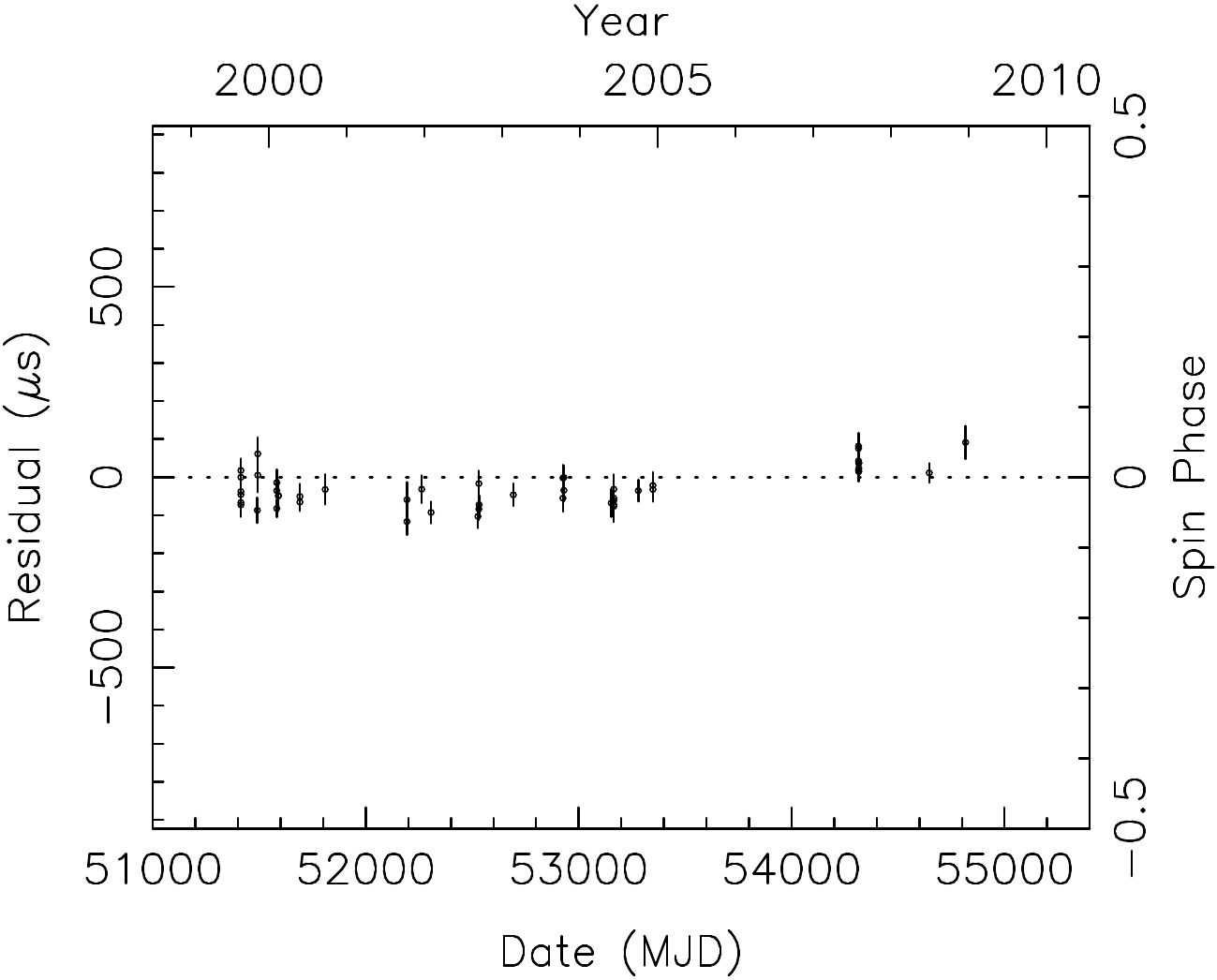}
    \caption{Pre-fit residuals using the ephemeris obtained with
    the second-best 10-gap solution.
    The narrow range of phases for all the TOAs outside these 10 gaps
    indicates that this particular ephemeris is
    based on the correct rotation count.}
    \label{fig:prediction}
\end{figure}

Using this script to look at all 10-gap solutions for 47~Tuc~aa, we find that
the ephemeris based on the second best (reduced $\chi^2\, = \, 1.37$,
integers presented in the last column of Table~\ref{tab:successive})
is able to predict all subsequent TOAs to less than 5\% of a rotation
(see Fig.~\ref{fig:prediction}). This behaviour is unique among
the 32 acceptable 10-gap solutions. This means that this is the correct solution.
Adding these integers to the rotation count predicted by the original
ephemeris we obtain the correct rotation count.

\begin{figure*}
	\includegraphics[width=\textwidth]{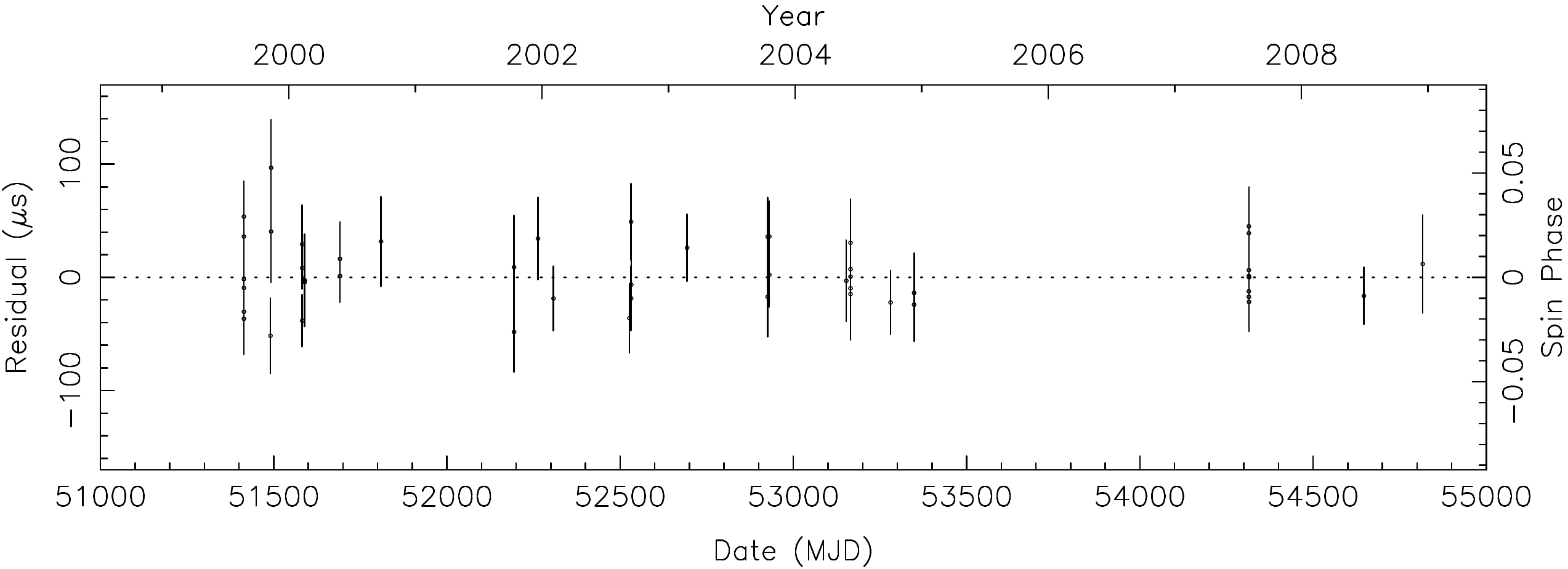}
    \caption{TOA residuals for 47 Tuc aa. Notice the sparseness of the data set.
    All residuals appear within a narrow range of
    spin phases. This happens without any JUMP statements, 
    indicating that the ephemeris is correct and can predict all times of arrival.
    No trends in the residuals
    are visible, this is an indication that the ephemeris gives a good description
    of the TOAs measured for the pulsar.
    \label{fig:residuals}}
\end{figure*}

We then use the ephemeris derived from the correct 10-gap solution
as the new starting ephemeris. This implies we must remove all previous PHASE
statements. Furthermore, because we're confident that this ephemeris
predicts the correct global rotation count, we can remove all 10 remaining gaps
(i.e., we remove all JUMP statements from
the remainder of the TOA.tim file) - if we were not yet confident about this,
we could use the same sort of test as in Section~\ref{sec:testing} to assure
ourselves that we indeed have unique ``0'' solutions for all remaining gaps.
With all gaps removed, we use {\sc tempo} to derive a global phase-coherent
ephemeris, with much improved parameters compared to the initial ephemeris.
This almost eliminates the
small residual trend we still see in Fig.~\ref{fig:prediction}, with
a post-fit reduced $\chi^2$ that is similar as in the previous step
(1.39), despite the fact that 10 degrees of freedom (the JUMP statements)
have been eliminated for this fit. This is another strong indication that
this is the correct solution. Fitting for $\mu_{\alpha}$, $\mu_{\delta}$ and 
the second spin frequency derivative ($\ddot{\nu}$), we obtain an even lower reduced
$\chi^2$, 1.33. The resulting ephemeris (with the EFAC
updated to 1.367 so as to yield a reduced $\chi^2$ of 1.0
and more conservative uncertainty estimates)
is presented in Table~\ref{tab:solution}.
{ The parameter uncertainties were derived using a
bootstrap method, similar to what is used by \cite{frk+17}.}
The TOA residuals calculated with this are depicted in Fig.~\ref{fig:residuals}.
The comparison of some of the parameters with those of the early partial
ephemerides is presented in Table~\ref{tab:parameters}.

\subsection{Timing parameters}

\begin{table}
\begin{center}{\scriptsize
\caption{Timing solution for PSR J0024$-$7205aa\label{tab:solution}. The parallax used is derived from the distance to 47~Tuc; it is not fitted.}
\begin{tabular}{l c}
\hline
\multicolumn{2}{c}{Observation and data reduction parameters}\\
\hline
  Fitting program \dotfill                                     & TEMPO  \\
  Time Units \dotfill                                          & TDB   \\
  Solar system ephemeris \dotfill                              & DE421  \\
  Reference Epoch (MJD) \dotfill                               & 51600   \\
  Span of Timing Data (MJD) \dotfill                           & 51143 - 54816  \\
  Number of TOAs \dotfill                                      &  49 \\
  EFAC \dotfill                                                & 1.367 \\
  RMS Residual ($\mu s$) \dotfill                              & 26.4 \\
  Reduced $\chi^2$ \dotfill                                    & 1.00 \\
\hline
\multicolumn{2}{c}{Timing parameters}\\
\hline  
  Right Ascension, $\alpha$ (J2000) \dotfill                   & 00:24:07.2783(8) \\
  Declination, $\delta$ (J2000) \dotfill                       & $-$72:05:19.5212(35) \\
  Spin frequency, $\nu$ (Hz)  \dotfill                       &  541.89365494899(14) \\
  First derivative of $\nu$, $\dot{\nu}$ ($10^{-14}\, \rm Hz\, s^{-1}$) \dotfill & +1.34754(35) \\
  Second derivative of $\nu$, $\ddot{\nu}$ ($10^{-26}\, \rm Hz \, s^{-2}$) \dotfill & +7.0(2.9) \\
  Proper motion in $\alpha$, $\mu_{\alpha}$ (mas yr$^{-1}$) \dotfill                   & 4.6(8)  \\
  Proper motion in $\delta$, $\mu_{\delta}$ (mas yr$^{-1}$) \dotfill                   & $-$4.6(1.3)  \\
  Parallax (mas) \dotfill & 0.2132 \\
  Dispersion Measure, DM (pc $\rm cm^{-3}$ ) \dotfill              & 24.941(7)   \\
\hline
\multicolumn{2}{c}{Derived parameters}\\
\hline
  Spin Period, $P$ (ms) \dotfill                            &  1.8453805296800(6) \\
  Spin Period Derivative, $\dot{P}$ ($10^{-20}$\,s\,s$^{-1}$) \dotfill  & $-$4.5890(15) \\
  Angular offset from centre in $\alpha$, $\theta_{\alpha}$ (\arcmin) \dotfill & 0.123 \\
  Angular offset from centre in $\delta$, $\theta_{\delta}$ (\arcmin) \dotfill & $-$0.448 \\
  Total angular offset from centre, $\theta_{\perp}$ (\arcmin) \dotfill & 0.465 \\
  $\theta_{\perp}$ (cluster core radii) \dotfill & 1.34 \\
  Projected offset from centre, $R_{\perp}$ (pc) \dotfill & 0.63 \\
\hline
\end{tabular}
\vspace{-0.5cm}
}
\end{center}
\end{table}

The pulsar ephemeris includes a very precise position.
As shown in Fig. 4 of \cite{frk+17}, the pulsar is located, in projection,
almost due South from the centre of the cluster, at a distance of 0.465 
arcminutes, or 1.34 core radii. 

Unlike virtually all pulsars in the Galactic disk, this pulsar
(and many others in globular clusters) has a negative $\dot{P}$.
For a rotationally powered pulsar, the intrinsic $\dot{P}$ is always positive;
this means that, for 47~Tuc~aa, this quantity is dominated
by the contribution from a negative acceleration of the pulsar
in the gravitational field of 47~Tuc (see discussion in \citealt{frk+17}). 
A negative acceleration means that the line-of-sight
acceleration of the pulsar is in the direction of the Earth; this
implies in turn that the pulsar is in the far half of the cluster.
This is expected from the relatively high DM of 47~Tuc~aa: previous 
observations have shown that the pulsars in the far side of the cluster 
(those with negative $\dot{P}$) have
larger DMs than the remaining pulsars. This happens because there is a cloud of ionized
gas in the centre of this globular cluster \citep{fkl+01}.
The magnitude of this negative acceleration can be explained by the model of the
gravitational field of the cluster presented by \cite{frk+17}, see Fig. 6 of that paper.

We also have a marginal detection of $\ddot{\nu}$;
this is caused by a change in the line-of-sight acceleration
in the gravitational field of the cluster that results from
the change of the position of the pulsar in the cluster;
these can also be caused by the motion relative to nearby stars.
The derived jerk is displayed graphically
with the jerks of other pulsars in Fig. 7 of \cite{frk+17}.

Since the TOA data set covers a total of 9 years, we can also measure the proper motion;
this is also listed in Table~\ref{tab:solution}. This proper motion is roughly consistent
with the proper motion of the remaining pulsars presented by \cite{frk+17}, but it has
a lower precision given the small number of TOAs.  

The precise location provided by the phase-coherent ephemeris
has allowed a detection of the pulsar in X-rays \citep{bhs+17}, as for the other 22 MSPs
in 47~Tuc with such ephemerides. 
Its X-ray flux is the smallest for any MSP in this cluster. This somewhat surprising find
is used to put an upper bound for amplitude of r-mode oscillations in this pulsar as
$\alpha \, < \, 2.5 \, \times \, 10^{-9}$ and to constrain the shape of the r-mode instability
window.

\section{Conclusions}
\label{sec:conclusions}

In the first part of this paper we have described a simple technique for determining the correct
global rotation count of any pulsar. This technique allows for a simple determination of
phase-coherent ephemerides for the vast majority of pulsars.
However, for some pulsars the sparseness and/or low precision of the TOAs
does not allow such a simple derivation of the global rotation count.

To solve such cases, we developed an algorithm that tests solutions for
$k$ gaps at the same time. Assuming each of the
acceptable solutions for $k - 1$ gaps, the algorithm searches (using
the ``gap mapper'') for acceptable solutions for gap number $k$; if these exist
then together with the assumed $(k-1)$-gap solution they make new $k$-gap solution(s).
$k$ is increased until a single $k$-gap solution becomes clearly superior to
the rest. The algorithm naturally eliminates solutions with a bad
reduced $\chi^2$ at every step. This is what we call ``sieving''.

This should not be described as a ``brute-force'' technique since estimating
the suitability of all combinations of acceptable 1-gap solutions
would clearly be impractical.
Sieving makes the problem not only tractable, but relatively cheap from a computational
point of view. Implementing the hierarchical parallel mode described in Section~\ref{sec:parallel}
and adding automatization should further reduce the amount of time needed to find
the correct rotation count with this method; this will likely be necessary for binary pulsars
with sparse data sets and/or poor timing precision. We have also suggested
changes to {\sc tempo} that might also produce further gains in speed and reliability; these issues have already been addressed in the development of
{\sc PINT}, which will likely be a better vehicle for future development.

We have implemented this technique with a simple shell script, which we have made freely available.
We have demonstrated the technique (and its implementation) by finding the
correct global rotation count and resultant phase-coherent ephemeris for
47~Tuc~aa, an isolated MSP in the globular cluster 47~Tuc with a very sparse set of
detections. The scientific implications of the parameters in this
ephemeris have already been presented in previous studies.

The automatic determination of the phase-coherent ephemeris for a MSP with such a
sparse set of detections implies that this process might be achievable with significantly
smaller amounts of telescope time than necessary until now. This will likely become a
more pressing issue once high-sensitivity instruments like the SKA come online:
this instrument will be finding many thousands of faint pulsars
\citep{2017arXiv171101910K,2017arXiv171201008L} that no other
telescope will be able to detect, which means that SKA time for follow-up will be
limited for each pulsar.

\section*{Acknowledgements}

We would like to thank our collaborators in the 47 Tuc timing project
(Michael Kramer, Dick Manchester, Andrew Lyne, Fernando Camilo, Dunc Lorimer,
Christine Jordan, John Sarkissian and Nichi D'Amico) for collecting and
helping to analyze the amazing AFB
data set on 47 Tuc over the years. We would also like to thank Norbert Wex and 
Andrew Cameron for suggestions and comments on the script, and Erik Madsen for 
his contributions to the Github implementation.
Both authors gratefully acknowledge financial support by the
European Research Council for the ERC  Starting grant BEACON under contract
No. 279702, and continued support from the Max Planck Society. Finally, we
acknowledge the referee, Dr. Scott Ransom, for valuable suggestions that have
significantly improved this work.











\bsp	
\label{lastpage}
\end{document}